\documentclass[aps,pre,twocolumn]{revtex4-1}
\pdfoutput=1

\usepackage{cmap}
\usepackage{graphicx}
\usepackage{fncylab,environ}
\usepackage{url}

\usepackage{amssymb}
\usepackage{amsmath}

\usepackage{tikz}
\usepackage{pgfplots}

\usepackage{hyperref}

\usetikzlibrary{external}
\newcommand{\scalarf}{1.}

\labelformat{figure}{Figure #1}
\labelformat{equation}{(#1)}
\labelformat{section}{Section #1}
\labelformat{table}{Table #1}

\pgfplotscreateplotcyclelist{list2}{
    {color=red},
    {color=brown},
    {color=violet},
    {color=green!60!black},
    {color=blue},
    {color=gray},
    {color=magenta},
    {color=orange},
    {color=olive}}

\newcommand\figgrzpfz{
\begin{figure}
    \centering
    \tikzsetnextfilename{grz-pfz}
    \scalebox{\scalarf}{
    \begin{tikzpicture}
	\begin{loglogaxis}[
		xlabel={$n$}, ylabel={$\langle f_\text{min}\rangle$},
		xtick = {200,400,800,1600}, xticklabels = {$200$,$400$,$800$,$1600$},
		xmin=180., xmax=2500.,
		ymin=0.001, ymax=0.1,
		axis y line = left,
		every axis y label/.style={rotate=90,at={(-0.05,0.85)}},
		legend entries={$\langle f_\text{min}\rangle$,$\langle r_\text{min}\rangle$},
		domain=180:2500,
	    ]
	    \addlegendimage{color=black, mark=diamond*, mark size=1.8pt};
	    \addlegendimage{color=black, mark=square*, mark size=1.2pt,dashed};
	    \addplot[only marks, mark=diamond*, mark size=0.7pt, error bars/y dir=both,error bars/y explicit] table[x index=0,y index=1, y error index=2] {fmin-slp.dat};
	    \addplot[color=black] {exp(0.3353988720789531*ln(10.) - 0.6893736143815803*ln(x)) };
	\end{loglogaxis}
	\begin{loglogaxis}[
		ylabel={$\langle r_\text{min}\rangle$},
		yticklabel pos = right,
		xtick = {200,400,800,1600}, xticklabels = {$200$,$400$,$800$,$1600$},
		axis y line = right,
		xmin=180., xmax=2500.,
		ymin=0.000005, ymax=0.0005,
		every axis y label/.style={rotate=90,at={(1.05,0.85)}},
		domain=180:2500,
	    ]
	    \addplot[only marks, mark=square*, mark size=0.4pt, error bars/y dir=both,error bars/y explicit] table[x index=0,y index=1, y error index=2] {rmin-slp.dat};
	    \addplot[color=black,dashed] {exp(0.5478976593924151*ln(10.) - 1.6787138693739025*ln(x)) };
	\end{loglogaxis}
    \end{tikzpicture}
    }
    \caption{\label{fig:grz-pfz}
	Mean values of the smallest force (solid line) and of the smallest slack (dashed line) in the locally
	optimal solutions reached by the sequential LP protocol in random instances of the perceptron as a function
	of system size. The lines show the power-law fit to the data, from which we extract the numerical
	exponent estimates: $\theta=0.451\pm0.006$ and $\gamma=0.404\pm0.004$.
    }
\end{figure}
}
\newcommand\figcrx{
    \begin{figure}
	\centering
	\tikzsetnextfilename{crx}
    \scalebox{\scalarf}{
	\begin{tikzpicture}
	    \begin{axis}[
		    xlabel = {$1/n$}, ylabel = {$\theta(n),\gamma(n)$},
		    scaled x ticks = false,
		    xtick = {0.001,0.002,0.003,0.004,0.005}, xticklabels = {$0.001$,$0.002$,$0.003$,$0.004$,$0.005$},
		    xmin = 0, xmax = 0.006,
		    ymin = 0.39, ymax = 0.49,
		    legend entries = {$\theta(n)$,$\gamma(n)$},
		    domain = 0:0.006,
		]
		\addlegendimage{color=black, mark=diamond*, mark size=1.8pt};
		\addlegendimage{color=black, mark=square*, mark size=1.8pt,dashed};
		\addplot[only marks, mark=diamond*, mark size=1.6pt, error bars/y dir=both,error bars/y explicit] table[x index=0,y index=1, y error index=2] {fmin-crx.dat};
		\addplot[only marks, mark=square*, mark size=1.6pt, error bars/y dir=both,error bars/y explicit] table[x index=0,y index=1, y error index=2] {rmin-crx.dat};
		\addplot[color=black] {0.42311};
		\addplot[color=black,dashed] {0.41269};
	    \end{axis}
	\end{tikzpicture}
    }
	\caption{\label{fig:crx}Critical exponent values inferred from the average slopes of the curves of \ref{fig:grz-pfz} measured between between $n$ and $2n$.
	The horizontal lines are the theoretically predicted values.
	No clear trend signaling corrections to finite-size scaling is apparent.}
    \end{figure}
}

\newcommand\figus{
\begin{figure*}
    \centering
    \scalebox{0.8}{
    \tikzsetnextfilename{scl-gr}
    \begin{tikzpicture}
	\begin{loglogaxis}[
		xlabel={$r$}, ylabel={$g(r)$},
		xmin=0.0000005, xmax=0.5,
		ymin=2., ymax=200.,
		every axis y label/.style={rotate=90,at={(-.05,0.5)}},
		cycle list name=list2,
	        only marks, mark size=0.4pt, error bars/y dir=both,error bars/y explicit,
		legend style={at={(0.97,0.97)},anchor=north east, font=\small},
		legend entries={$200$,$300$,$400$,$600$,$800$,$1200$,$1600$,$2400$}
	    ]
	    \addplot table[x index=0,y index=1, y error index=2] {uns-gr-slp-1.dat};
	    \addplot table[x index=0,y index=1, y error index=2] {uns-gr-slp-2.dat};
	    \addplot table[x index=0,y index=1, y error index=2] {uns-gr-slp-3.dat};
	    \addplot table[x index=0,y index=1, y error index=2] {uns-gr-slp-4.dat};
	    \addplot table[x index=0,y index=1, y error index=2] {uns-gr-slp-5.dat};
	    \addplot table[x index=0,y index=1, y error index=2] {uns-gr-slp-6.dat};
	    \addplot table[x index=0,y index=1, y error index=2] {uns-gr-slp-7.dat};
	    \addplot table[x index=0,y index=1, y error index=2] {uns-gr-slp-8.dat};
	\end{loglogaxis}
	\begin{loglogaxis}[
		xshift={0.382\linewidth}, ylabel={}, yticklabels={},
		xlabel={$r n^{1/(1-\gamma)}$}, 
		xmin=0.02, xmax=100000.,
		ymin=0.01, ymax=1.,
		cycle list name=list2,
	        only marks, mark size=0.4pt, error bars/y dir=both,error bars/y explicit,
	    ]
	    \addplot table[x index=0,y index=1, y error index=2] {scl-gr-slp-1.dat};
	    \addplot table[x index=0,y index=1, y error index=2] {scl-gr-slp-2.dat};
	    \addplot table[x index=0,y index=1, y error index=2] {scl-gr-slp-3.dat};
	    \addplot table[x index=0,y index=1, y error index=2] {scl-gr-slp-4.dat};
	    \addplot table[x index=0,y index=1, y error index=2] {scl-gr-slp-5.dat};
	    \addplot table[x index=0,y index=1, y error index=2] {scl-gr-slp-6.dat};
	    \addplot table[x index=0,y index=1, y error index=2] {scl-gr-slp-7.dat};
	    \addplot table[x index=0,y index=1, y error index=2] {scl-gr-slp-8.dat};
	\end{loglogaxis}
	\begin{loglogaxis}[
		xshift={0.764\linewidth},
		xlabel={$r n^{1/(1-\gamma)}$}, ylabel={$g(r) n^{-\gamma/(1-\gamma)}$},
		every axis y label/.style={rotate=90,at={(1.05,0.25)}},
		xmin=0.02, xmax=100000., yticklabel pos = right,
		ymin=0.01, ymax=1.,
		cycle list name=list2,
	        only marks, mark size=0.4pt, error bars/y dir=both,error bars/y explicit,
	    ]
	    \addplot table[x index=0,y index=1, y error index=2] {ths-gr-slp-1.dat};
	    \addplot table[x index=0,y index=1, y error index=2] {ths-gr-slp-2.dat};
	    \addplot table[x index=0,y index=1, y error index=2] {ths-gr-slp-3.dat};
	    \addplot table[x index=0,y index=1, y error index=2] {ths-gr-slp-4.dat};
	    \addplot table[x index=0,y index=1, y error index=2] {ths-gr-slp-5.dat};
	    \addplot table[x index=0,y index=1, y error index=2] {ths-gr-slp-6.dat};
	    \addplot table[x index=0,y index=1, y error index=2] {ths-gr-slp-7.dat};
	    \addplot table[x index=0,y index=1, y error index=2] {ths-gr-slp-8.dat};
	\end{loglogaxis}
    \end{tikzpicture}
    }\\\scalebox{0.8}{
    \tikzsetnextfilename{scl-pf}
    \begin{tikzpicture}
	\begin{loglogaxis}[
		xlabel={$f$}, ylabel={$p(f)$},
		xmin=0.0003, xmax=10,
		ymin=0.02, ymax=1.,
		cycle list name=list2,
	        only marks, mark size=0.4pt, error bars/y dir=both,error bars/y explicit,
		every axis y label/.style={rotate=90,at={(-.05,0.65)}},
		legend style={at={(0.87,0.03)},anchor=south east, font=\small},
		legend entries={$200$,$300$,$400$,$600$,$800$,$1200$,$1600$,$2400$}
	    ]
	    \addplot table[x index=0,y index=1, y error index=2] {uns-pf-slp-1.dat};
	    \addplot table[x index=0,y index=1, y error index=2] {uns-pf-slp-2.dat};
	    \addplot table[x index=0,y index=1, y error index=2] {uns-pf-slp-3.dat};
	    \addplot table[x index=0,y index=1, y error index=2] {uns-pf-slp-4.dat};
	    \addplot table[x index=0,y index=1, y error index=2] {uns-pf-slp-5.dat};
	    \addplot table[x index=0,y index=1, y error index=2] {uns-pf-slp-6.dat};
	    \addplot table[x index=0,y index=1, y error index=2] {uns-pf-slp-7.dat};
	    \addplot table[x index=0,y index=1, y error index=2] {uns-pf-slp-8.dat};
	\end{loglogaxis}
	\begin{loglogaxis}[
		xshift={0.382\linewidth},
		xlabel={$f n^{1/(1+\theta)}$}, ylabel={}, yticklabels={},
		xmin=0.03, xmax=1000,
		ymin=0.1, ymax=10.,
		cycle list name=list2,
	        only marks, mark size=0.4pt, error bars/y dir=both,error bars/y explicit,
	    ]
	    \addplot table[x index=0,y index=1, y error index=2] {scl-pf-slp-1.dat};
	    \addplot table[x index=0,y index=1, y error index=2] {scl-pf-slp-2.dat};
	    \addplot table[x index=0,y index=1, y error index=2] {scl-pf-slp-3.dat};
	    \addplot table[x index=0,y index=1, y error index=2] {scl-pf-slp-4.dat};
	    \addplot table[x index=0,y index=1, y error index=2] {scl-pf-slp-5.dat};
	    \addplot table[x index=0,y index=1, y error index=2] {scl-pf-slp-6.dat};
	    \addplot table[x index=0,y index=1, y error index=2] {scl-pf-slp-7.dat};
	    \addplot table[x index=0,y index=1, y error index=2] {scl-pf-slp-8.dat};
	\end{loglogaxis}
	\begin{loglogaxis}[
		xshift={0.764\linewidth}, ylabel={$p(f) n^{\theta/(1+\theta)}$},
		xlabel={$f n^{1/(1+\theta)}$},
		xmin=0.03, xmax=1000, yticklabel pos = right,
		ymin=0.1, ymax=10.,
		cycle list name=list2,
		every axis y label/.style={rotate=90,at={(1.05,0.25)}},
	        only marks, mark size=0.4pt, error bars/y dir=both,error bars/y explicit,
	    ]
	    \addplot table[x index=0,y index=1, y error index=2] {ths-pf-slp-1.dat};
	    \addplot table[x index=0,y index=1, y error index=2] {ths-pf-slp-2.dat};
	    \addplot table[x index=0,y index=1, y error index=2] {ths-pf-slp-3.dat};
	    \addplot table[x index=0,y index=1, y error index=2] {ths-pf-slp-4.dat};
	    \addplot table[x index=0,y index=1, y error index=2] {ths-pf-slp-5.dat};
	    \addplot table[x index=0,y index=1, y error index=2] {ths-pf-slp-6.dat};
	    \addplot table[x index=0,y index=1, y error index=2] {ths-pf-slp-7.dat};
	    \addplot table[x index=0,y index=1, y error index=2] {ths-pf-slp-8.dat};
	\end{loglogaxis}
    \end{tikzpicture}
    }
    \caption{\label{fig:uns}\label{fig:scl}
        Empirical distributions (left) of slacks of inactive constraints (top) and Lagrange multipliers of active constraints (bottom)
	at the jamming point. The middle and right panels show the finite-size scaling collapse of the distributions using the numerically extracted exponents
	and the theoretically predicted exponents respectively.
    }
\end{figure*}
}

\newcommand\figsclecej{
\begin{figure}
    \centering
    \scalebox{\scalarf}{
    \tikzsetnextfilename{sclecej}
    \begin{tikzpicture}
	\begin{loglogaxis}[
		xlabel={$fn^{1/(1+\theta)}$}, ylabel={$p(f)n^{\theta/(1-\theta)}$},
		xmin=0.0008, xmax=1000,	ymin=0.03, ymax=8,
		cycle list name=list2,
		only marks, mark size=0.4pt, error bars/y dir=both,error bars/y explicit,
		legend style={at={(0.72,0.03)},anchor=south east, font=\tiny},
		legend entries={$30$,$45$,$60$,$90$,$120$,$180$,$240$,$360$,$480$},
	    ]
	    \addplot table[x index=0,y index=1, y error index=2] {scaled-pf-hc-n1.dat};
	    \addplot table[x index=0,y index=1, y error index=2] {scaled-pf-hc-n2.dat};
	    \addplot table[x index=0,y index=1, y error index=2] {scaled-pf-hc-n3.dat};
	    \addplot table[x index=0,y index=1, y error index=2] {scaled-pf-hc-n4.dat};
	    \addplot table[x index=0,y index=1, y error index=2] {scaled-pf-hc-n5.dat};
	    \addplot table[x index=0,y index=1, y error index=2] {scaled-pf-hc-n6.dat};
	    \addplot table[x index=0,y index=1, y error index=2] {scaled-pf-hc-n7.dat};
	    \addplot table[x index=0,y index=1, y error index=2] {scaled-pf-hc-n8.dat};
	    \addplot table[x index=0,y index=1, y error index=2] {scaled-pf-hc-n9.dat};
	    \addplot table[x index=0,y index=1, y error index=2] {scaled-pf-ej-n1.dat};
	    \addplot table[x index=0,y index=1, y error index=2] {scaled-pf-ej-n2.dat};
	    \addplot table[x index=0,y index=1, y error index=2] {scaled-pf-ej-n3.dat};
	    \addplot table[x index=0,y index=1, y error index=2] {scaled-pf-ej-n4.dat};
	    \addplot table[x index=0,y index=1, y error index=2] {scaled-pf-ej-n5.dat};
	    \addplot table[x index=0,y index=1, y error index=2] {scaled-pf-ej-n6.dat};
	    \addplot table[x index=0,y index=1, y error index=2] {scaled-pf-ej-n7.dat};
	    \addplot table[x index=0,y index=1, y error index=2] {scaled-pf-ej-n8.dat};
	    \addplot table[x index=0,y index=1, y error index=2] {scaled-pf-ej-n9.dat};
	\end{loglogaxis}
    \end{tikzpicture}
    }
    \caption{\label{fig:sclecej}
        Finite-size scaling collapse for jamming points of the edge-climbing and edge-jumping protocols.
	The two sets of data collapse to a common curve
	in the critical regime, but to two separate scaling curves in the subcritical regime.
    }
\end{figure}
}

\makeatletter

\begin{document}

\title{Scaling collapse at the jamming transition}

\author{Yoav Kallus}
\affiliation{Santa Fe Institute, 1399 Hyde Park Road, Santa Fe, New Mexico 87501}

\date{\today}

\begin{abstract}
    The jamming transition of particles with finite-range interactions
    is characterized by a variety of critical phenomena, including power
    law distributions of marginal contacts. We numerically study
    a recently proposed simple model of jamming, which is conjectured to
    lie in the same universality class as the jamming of spheres
    in all dimensions. We extract numerical estimates of the critical exponents,
    $\theta=0.451\pm0.006$ and $\gamma=0.404\pm0.004$,
    that match the exponents observed in sphere packing systems. We analyze
    finite-size scaling effects that manifest in a subcritical cutoff regime
    and size-independent, but protocol-dependent scaling curves. Our results
    supports the conjectured link with sphere jamming, provide more precise
    measurements of the critical exponents than previously reported,
    and shed light on the finite-size scaling behavior of continuous constraint
    satisfiability transitions.
\end{abstract}
\pacs{05.65.+b,45.70.-n,89.75.Da,89.75.Fb}

\maketitle

Jamming is the continuous emergence of nonzero
mechanical moduli in a system of particles with finite-range interactions \cite{jam1,jam2,jam3}.
The ensemble of configurations
at the transition boundary exhibits many dramatic
phenomena, including anomalous vibration spectra \cite{softmodes2,softmodes1,fpuz},
isostaticity and hyperuniformity \cite{hyperuniform}, critical scaling of mechanical moduli \cite{mechscale},
and power-law distributions of the weakest forces and the smallest
interparticle gaps \cite{jam-struct,fp,iel1}. The jamming transition can be understood as a
continuous analog of the satisfiability transition in
random constraint satisfaction problems \cite{fp}, usually studied in the discrete setting \cite{sat3}.
Recently the transition in a simple constraint problem, the perceptron,
was conjectured to lie in the same universality class as the sphere jamming
transition in all spatial dimensions \cite{fp}. Here we numerically study the perceptron
and show that the power law exponents match those
measured numerically for sphere jamming, but may not match
the theoretically predicted exponents. We employ finite-size scaling
to extract precise measurements of the exponents and demonstrate
the universal critical behavior of the system. 
Our results support the conjecture linking the perceptron to sphere jamming,
and provide more precise numerical estimates of the jamming critical exponents
than previously reported. 

The jamming ensemble corresponds to the random packing configurations
of nonoverlapping hard spheres
such that no improvement in the packing density is possible by local
movement of the spheres. As such, it is a local transition
between the satisfiability of the nonoverlap constraints at lower density,
and unsatisfiabily of all constraints simultaneously at higher density.
This transition for equal-sized
three-dimensional spheres occurs typically at a range of densities around $64\%$
depending on preparation protocol \cite{Chaudhuri}. The restriction to random
packing is crucial, since the constraints are simultaneously satisfiable
all the way to the close packing density of spheres, $74\%$. As a result, any
theoretical treatment of the jamming transition must overcome the problem
of excluding configurations with crystalline order from the phase space.
One approach that has been successful in reproducing the metastable disordered phase
behavior of systems with a thermodynamically stable ordered phase has been
to construct systems with quenched disordered that exhibit the same correlation
structure in the disordered phase \cite{Marinari1994}.

The perceptron model, well known in machine learning \cite{perceptron}, when extended
to its nonconvex regime, has been shown by Franz and Parisi to be solved by the same
replica symmetry breaking (RSB) ansatz as that occurring as a solution for the jamming
point in a model of sphere packing in the limit of an infinite number of dimensions \cite{fp,sp-inf}.
It is conjectured that the jamming transitions of spheres in all dimensions lie in the
same universality class as these two models \cite{jam-struct,sp-inf}. 
Here, we numerically simulate the perceptron model and compare our numerical results to
numerical simulations of sphere jamming and the theoretical predictions of the RSB
solution.

For a given size $n$ and number of constraints $m$, an instance of the perceptron is given by the
following constrained optimization problem:
\begin{equation}\label{eq:FP-opt}
    \begin{aligned}
	\text{maximize }&F(\mathbf{x})=\|\mathbf{x}\|^2=\langle\mathbf{x},\mathbf{x}\rangle\\
	\text{subject to }&\mathbf{x}\in \mathbb{R}^n\\
	&\langle\mathbf{\xi}_\mu,\mathbf{x}\rangle\le1\text{ for all }\mu=1,\ldots,m\text,
    \end{aligned}
\end{equation}
where $\mathbf{\xi}_\mu$, $\mu=1,\ldots,m$, are independently drawn
from a uniform distribution over the unit sphere $S^{n-1}$. To approach the thermodynamic
limit, the system size $n$ is increased toward infinity while the ratio $\alpha=m/n$
is held constant \cite{fp}.

An equivalent formulation, closer to the traditional setting of the perceptron model
is obtained by restricting $\mathbf{x}$ to $S^{n-1}$ and maximizing
the distance from the nearest point $\xi_\mu$. However, the formulation \ref{eq:FP-opt}
allows all the constraints to be linear inequality constraints. Thus,
the only nonlinearity, distinguishing \ref{eq:FP-opt} from a linear program (LP), is in the
objective function. Moreover, since the objective is convex, the local
maxima occur at vertices of the feasible polytope, so the number of active
constraints (constraints satisfied with equality) must be at least $n$.
(It can also be no greater than $n$ due to the genericity of the constraints).
In these properties, the model is reminiscent of an earlier jamming model,
namely the jamming transition of lattice sphere packing \cite{iel1,iel2}. In this model,
the determinant of a symmetric $d\times d$ matrix $Q$ is minimized, subject to the linear
constraints $\langle \mathbf{\xi}, Q\mathbf{\xi}\rangle\ge1$ for all $\xi\in \mathbb{Z}^d$.
Whereas the perceptron model has quenched disorder, the sphere packing and lattice
packing models do not and have an ordered phase in addition to the disordered phase.

One of the remarkable properties of the jamming ensemble observed in sphere jamming is the power
law distribution of very weak contact forces and of very small interparticle
gaps. In the context of general constrained optimization problems, these correspond to the
distribution of Lagrange (or Karush--Kuhn--Tucker) multipliers associated with
active inequality constraints and the
distribution of slacks of inactive constraints (the difference between the two sides
of an inequality constraint).
The constraints at the low end of each of these distributions can be thought of
as the marginal constraints: the active constraints closest to becoming inactive and \textit{vice
versa}. The observation of power law distributions of marginal constraints has been
explained heuristically as resulting from the marginal stability of typical
jammed configuration \cite{Wyart2012}: in order for the configuration to be stable, it must either
exhibit a suppression of weak active constraints, which contribute to instability,
or an enhancement of small-slack inactive constraints, which contribute to stability,
or both, as is observed. A quantitative version of this argument gives a predicted
relation between the exponents of the two power laws.

\figgrzpfz

At a locally optimal feasible point of \ref{eq:FP-opt}, standard results in nonlinear
optimization imply the existence of Lagrange multipliers (forces) $f_\mu$,
such that $f_\mu>0$ if and only if $\langle\mathbf{\xi}_\mu, \mathbf{x}\rangle =1$, and
$\mathbf{x} = \|\mathbf{x}\| n^{-1/2} \sum_{\mu}f_\mu \mathbf{\xi}_\mu$.
It is predicted that in the thermodynamic limit, the distribution of forces satisfies
a power law $p(f)\sim f^\theta$ with a universal exponent $\theta=0.42311$ for small $f$ \cite{fp}. In particular,
the smallest force is of the order of the value $f_\text{min}$ such that $\int_0^{f_\text{min}} p(f)df\sim\tfrac1n$.
That is, $f_\text{min}\sim n^{-1/(1+\theta)}$. For finite systems, we expect the
distribution to follow the power law only in the range $n^{-1/(1+\theta)}\ll f\ll1$. However,
as in other universal critical points, we expect a finite-size scaling behavior
even for moderately sized finite systems. In particular, we expect, when $f\ll1$,
\begin{equation}\label{eq:scale-pf}
    p(f) = n^{-\theta/(1+\theta)} \tilde{p}\left(f n^{1/(1+\theta)}\right)\text.
\end{equation}

We are also interested in the distribution of slacks of inactive constraints.
Let $r_\mu = n^{1/2} (1-\langle\mathbf{\xi}_\mu,\mathbf{x}\rangle)/\|\mathbf{x}\|$
and let $n g(r)dr$ be the expected number of constraints $\mu$ such that
$0<r\le r_\mu\le r+dr$. The theoretically predicted distribution for small
$r$ in the thermodynamic limit is $g(r)\sim r^{-\gamma}$, and consequently
the smallest nonzero slack should scale with system size as $r_\text{min}\sim n^{-1/(1-\gamma)}$ \cite{fp}.
The theoretically predicted universal value for the exponent is $\gamma=0.41269$ \cite{fp}.
For finite systems, the power law should be cut off around $r_\text{min}$ consistent
with a finite-size scaling function
\begin{equation}\label{eq:scale-gr}
    g(r) = n^{\gamma/(1-\gamma)} \tilde{g}\left(r n^{1/(1-\gamma)}\right).
\end{equation}
The marginal stability heuristic predicts that $\gamma = 1/(2+\theta)$,
a relationship that is satisfied by the theoretically predicted values \cite{Wyart2012,fp}.

\figcrx

\figus

We perform numerical optimization of the perceptron at a constraint ratio $\alpha=4$
and sizes $n=200$, $300$, $400$, $600$, $800$, $1200$, $1600$, and $2400$. We generate
$10^5$, $10^5$, $10^5$, $10^5$, $10^5$, $5\cdot 10^4$, $2\cdot 10^4$, and $2\cdot 10^4$
independent realizations, respectively, of each size. The optimization protocol
solves a sequence of LPs that are identical to \ref{eq:FP-opt}, except that
the objective is replaced by a linear objective $F_t(\mathbf{x}) = \langle\mathbf{y}_t,\mathbf{x}\rangle$.
The objective vector for each step is given by the solution of the previous LP:
$\mathbf{y}_{t+1} = \mathbf{x}^\text{opt}_t$. The real objective $\|\mathbf{x}\|^2$
necessarily increases at each step and the process converges to a local
optimum of \ref{eq:FP-opt} in finitely many steps.

In \ref{fig:grz-pfz}, we plot the mean value of the smallest force $\langle f_\text{min}\rangle$
and the mean value of the smallest slack $\langle r_\text{min}\rangle$
as functions of the system size. In both cases, we obtain a power-law fit
as predicted. The inferred values of the critical exponents are
$\gamma=0.404\pm0.004$ and $\theta=0.451\pm0.006$. The inferred values are close
to the theoretically predicted values, but do not seem to include the
latter in their confidence regions. Still, the inferred values
satisfy the marginality relation $\gamma=1/(2+\theta)$.

It is possible that corrections to
finite-size scaling are responsible for the discrepancy between our numerically-inferred
exponents and the theoretical values and that the
apparent exponents at finite $n$ are different from their limit as $n\to\infty$.
We measure the local slopes of the curves of \ref{fig:grz-pfz}, and we
do not observe a clear trend in $\gamma(n)$ or $\theta(n)$,
which would be indicative of corrections to finite-size scaling (see \ref{fig:crx}).
Still, partly due to the increase in the error estimate, the slope values at
larger sizes seem to be more consistent with the theoretically
predicted exponents. Another explanation could be that the thermodynamic
calculation predicting the exponent values only approximately applies
to the setting at hand, which employs a highly off-equilibrium hill-climbing dynamic.

\figsclecej

Previous numerical studies of systems conjectured to be in the same universality
class provide similar estimates of the critical exponents. Studying hard and
soft spheres in up to 10 dimensions, the authors of Ref.\ \cite{jam-struct} find
$\gamma\approx0.39$ for soft spheres, and $\gamma\approx0.42$
for hard spheres. In finite-dimensional
sphere packings, the distribution of contact forces has additional contributions
due to localized modes that dominates at low forces, but the relevant
exponent $\theta_e$ can be extracted by looking only at contacts coupled to extended
modes \cite{Lerner2013,extended}. In Ref.\ \cite{Lerner2013}, the authors measure $\gamma\approx0.38$
and $\theta_e\approx0.44$ for hard spheres in 3 dimensions. 
In Ref.\ \cite{DeGiuli2014}, the authors measure $\gamma\approx 0.4$
and $\theta_e\approx0.44$ for hard spheres in 2 and 3 dimensions. These values
closely agree with our estimates of the critical exponents in the perceptron.
Since we approach the transition from the satisfiable side, our results should
be considered analogous to the hard sphere results.

We now turn to inspecting the distributions $g(r)$ and $p(f)$ presented in \ref{fig:uns}.
As the system size increases, a clear power law develops extending to lower and lower
values of $r$ and $f$. However, at each finite size, the power law is cut off and approaches
a constant as $r,f\to0$. The finite nonzero values of $g(0)$ and $p(0)$,
are of particular interest. Observation of such cutoffs have previously been cited as evidence against
the existence of a power law (or for trivial exponents $\gamma=\theta=0$) \cite{Donev2005,fnet1,fnet2,iel2,cavity},
but in the current context can be understood as finite-size effects.

The region controlled entirely by the power law appears to be no more than
two decades of magnitude at the largest system size for each distribution.
Therefore, even when we use the steepest decade of data at the largest system size
to fit a power law, we obtain exponent estimates significantly lower than
the ones inferred by comparing different system sizes: $\gamma\approx0.37$ and $\theta\approx0.42$.
The shallower behavior in the cutoff regimes affects the curve significantly
even dacades away from the overt crossover.
This appears to not be the case for sphere jamming,
where the sparsity of the constraints in comparison with the number of degrees
of freedom seems to allow the efficient simulation of much larger systems, where
the power-law regime spans 3--4 decades. Even in that case, the subcritical and
supercritical regimes could still have an effect on the apparent power-law
exponents. Studying the finite-size
scaling of the distribution could provide a useful comparison.

We can test the finite-size scaling hypotheses, \ref{eq:scale-pf} and \ref{eq:scale-gr},
by appropriately rescaling the axes by the appropriate powers of $n$. When we use our
inferred exponent values to rescale the data (see \ref{fig:scl}), they collapse to a
size-independent curve, with a common behavior in the critical and subcritical regimes.
The theoretically predicted exponent values give a slightly, but visibly, poorer collapse,
especially for the distribution of slacks.

While the scaling curve appears to be size independent, we do observe striking protocol dependence:
in addition to the sequential LP protocol described above, we implemented two additional protocols
at smaller sizes. The edge-climbing (EC) protocol mimics the simplex algorithm of linear programming,
starting at a vertex of the feasible region and traveling at each step to an adjacent vertex
along the edge offering the largest directional derivative of the objective. When all such derivatives
are negative, a local optimum is achieved and the protocol terminates. The edge-jumping (EJ) protocol
makes EC moves when they are available, but otherwise tries each edge and follows it if the objective
value at the other endpoint is larger. If neither type of move is available, the protocol terminates.

We see an overt difference in the distribution of forces at the jamming point achieved by the two
protocols (\ref{fig:sclecej}). While $p(f)$ shows similar behaviors in the power law regime, 
below the cutoff we observe $p(f)\sim\text{constant}$ in the EC protocol,
and $p(f)\sim f$ in the EJ protocol. The edge-jumping moves destabilize weak constraints and
suppress the subcritical regime of the force distribution in a nonuniversal way,
clearly distinguished from the universal critical suppression.

The perceptron model provides a numerically tractable setting in which to test
many of the theoretical ideas and methods about the jamming transition.
Finite-size effects studied here could be crucial to applying these ideas
to real-world problems away form the thermodynamic limit. In the present case at least,
comparison of systems of different sizes provides a more sensitive measure of
the critical exponents, which are distorted by finite-size effects in
any single size studied here. Our estimates of the critical exponents match those measured numerically in sphere
jamming systems, supporting the conjecture that they are in the same universality class.
However, we observe some variance from the theoretically predicted exponent values,
both in the scaling of the minimal force and minimal slack with system size,
and in the finite-size scaling collapse. Finally, we observe size-independent but
protocol-dependent finite-size scaling curves that demonstrate distinct critical
and subcritical behavior.

\textbf{Acknowledgment}. This work was supported by an Omidyar Fellowship at the Santa Fe Institute. 

\bibliographystyle{alpha}
\bibliography{con.bib}

\newcommand{\etalchar}[1]{$^{#1}$}
\begin{thebibliography}{TSVvH10}

\bibitem[BBB{\etalchar{+}}11]{jam1}
L.~Berthier, G.~Biroli, J.-P. Bouchaud, L.~Cipelletti, and W.~van Saarloos,
  editors.
\newblock {\em Dynamical Heterogeneities in Glasses, Colloids, and Granular
  Media}.
\newblock Oxford Univ Press, Oxford, 2011.

\bibitem[BMSM14]{cavity}
L.~Bo, R.~Mari, C.~Song, and H.~A. Makse.
\newblock Cavity method for force transmission in jammed disordered packings of
  hard particles.
\newblock {\em Soft Matter}, 10:7379--7392, 2014.

\bibitem[CBS10]{Chaudhuri}
P.~Chaudhuri, L.~Berthier, and S.~Sastry.
\newblock Jamming transitions in amorphous packings of frictionless spheres
  occur over a continuous range of volume fractions.
\newblock {\em Phys. Rev. Lett.}, 104:165701, 2010.

\bibitem[CCPZ12]{jam-struct}
P.~Charbonneau, E.~I. Corwin, G.~Parisi, and F.~Zamponi.
\newblock Universal microstructure and mechanical stability of jammed packings.
\newblock {\em Phys. Rev. Lett.}, 109(20), 2012.

\bibitem[CCPZ15]{extended}
P~Charbonneau, E.~I. Corwin, G.~Parisi, and F.~Zamponi.
\newblock Jamming criticality revealed by removing localized buckling
  excitations.
\newblock {\em Phys. Rev. Lett.}, 114:125504, 2015.

\bibitem[CKP{\etalchar{+}}14]{sp-inf}
P.~Charbonneau, J.~Kurchan, G.~Parisi, P.~Urbani, and F.~Zamponi.
\newblock Exact theory of dense amorphous hard spheres in high dimension {III}:
  The full replica symmetry breaking solution.
\newblock {\em J. Stat. Mech.}, 2014(10):P10009, 2014.

\bibitem[DLBW14]{DeGiuli2014}
E.~DeGiuli, E.~Lerner, C.~Brito, and M.~Wyart.
\newblock Force distribution affects vibrational properties in hard-sphere
  glasses.
\newblock {\em Proc. Natl. Acad. Sci. USA}, 111(48):17054--17059, 2014.

\bibitem[DTS05]{Donev2005}
A.~Donev, S.~Torquato, and F.~H. Stillinger.
\newblock Pair correlation function characteristics of nearly jammed disordered
  and ordered hard-sphere packings.
\newblock {\em Phys. Rev. E}, 71:011105, 2005.

\bibitem[FP15]{fp}
S.~Franz and G.~Parisi.
\newblock The simplest model of jamming.
\newblock arXiv:1501.03397, 2015.

\bibitem[FPUZ15]{fpuz}
F.~Franz, G.~Parisi, P.~Urbani, and F.~Zamponi.
\newblock Universal spectrum of normal modes in low-temperature glasses: an
  exact solution.
\newblock {\em Proc. Nat. Acad. Sci. USA}, 112(47):14539–14544, 2015.

\bibitem[KMT13]{iel1}
Y.~Kallus, {\'E.}~Marcotte, and S.~Torquato.
\newblock Jammed lattice sphere packings.
\newblock {\em Phys. Rev. E}, 88(6):062151, 2013.

\bibitem[KT14]{iel2}
Y.~Kallus and S.~Torquato.
\newblock Marginal stability in jammed packings: Quasicontacts and weak
  contacts.
\newblock {\em Phys. Rev. E}, 90(2):022114, 2014.

\bibitem[LDW13]{Lerner2013}
E.~Lerner, G.~{D\"uring}, and M.~Wyart.
\newblock Low-energy non-linear excitations in sphere packings.
\newblock {\em Soft Matter}, 9:8252--8263, 2013.

\bibitem[MPR94]{Marinari1994}
E.~Marinari, G.~Parisi, and F.~Ritort.
\newblock Replica field theory for deterministic models: {II.} {A} non-random
  spin glass with glassy behavior.
\newblock {\em J. Phys. A: Math. Gen.}, 27:7646--7668, 1994.

\bibitem[OLLN02]{jam3}
C.~S. O'Hern, S.~A. Langer, A.~J. Liu, and S.~R. Nagel.
\newblock Random packings of frictionless particles.
\newblock {\em Physical Review Letters}, 88(7), 2002.

\bibitem[Par14]{softmodes1}
G.~Parisi.
\newblock Soft modes in jammed hard spheres {(I)}: {Mean} field theory of the
  isostatic transition.
\newblock 2014.
\newblock arXiv:1403.4413.

\bibitem[PIM06]{sat3}
A.~Percus, G.~Istrate, and C.~Moore, editors.
\newblock {\em Computational Complexity and Statistical Physics}.
\newblock Oxford Univ Press, Oxford, 2006.

\bibitem[Ros58]{perceptron}
F.~Rosenblatt.
\newblock The perceptron: A probabilistic model for information storage and
  organization in the brain.
\newblock {\em Psychological Review}, 65(6):386--408, 1958.

\bibitem[TS10]{jam2}
S.~Torquato and F.~H. Stillinger.
\newblock Jammed hard-particle packings: From kepler to bernal and beyond.
\newblock {\em Rev. Mod. Phys.}, 82:2633, 2010.

\bibitem[TSS{\etalchar{+}}05]{fnet1}
B.~P. Tighe, J.~E.~S. Socolar, D.~G. Schaeffer, W.~G. Mitchener, and M.~L.
  Huber.
\newblock Force distributions in a triangular lattice of rigid bars.
\newblock {\em Phys. Rev. E}, 72:031306, 2005.

\bibitem[TSVvH10]{fnet2}
B.~P. Tighe, J.~H. Snoeijer, T.~J.~H. Vlugt, and M.~van Hecke.
\newblock The force network ensemble for granular packings.
\newblock {\em Soft Matter}, 6:2908--2917, 2010.

\bibitem[vH09]{mechscale}
M.~van Hecke.
\newblock Jamming of soft particles: geometry, mechanics, scaling and
  isostaticity.
\newblock {\em J. Phys.: Cond. Matt.}, 22(3):033101, 2009.

\bibitem[WNW05]{softmodes2}
M.~Wyart, S.~R. Nagel, and T.~A. Witten.
\newblock Geometric origin of excess low-frequency vibrational modes in weakly
  connected amorphous solids.
\newblock {\em Europhys. Lett.}, 72(3):486--492, 2005.

\bibitem[Wya12]{Wyart2012}
M.~Wyart.
\newblock Marginal stability constrains force and pair distributions at random
  close packing.
\newblock {\em Phys. Rev. Lett.}, 109(12):125502, 2012.

\bibitem[ZJT11]{hyperuniform}
C.~E. Zachary, Y.~Jiao, and S.~Torquato.
\newblock Hyperuniform long-range correlations are a signature of disordered
  jammed hard-particle packings.
\newblock {\em Phys. Rev. Lett.}, 106:178001, 2011.

\end{thebibliography}
\end{document}